\documentclass[twocolumn,floatfix]{revtex4}
\RequirePackage[colorlinks,hyperindex]{hyperref}
\usepackage{dcolumn}
\usepackage{bm}

\usepackage{graphicx}
\begin{document}

\title{Acoustoelectric luminescence from a field-effect \textit{n-i-p} lateral junction}

\author{Giorgio De Simoni\footnote{electronic mail: g.desimoni@sns.it}}
\author{Vincenzo Piazza}
\author{Lucia Sorba\footnote{Also at Laboratorio Nazionale TASC, CNR-INFM, AREA Science Park, Trieste I-34012 Trieste, Italy}}
\affiliation{NEST, Scuola Normale Superiore and CNR-INFM, Piazza dei Cavalieri 7, I-56126 Pisa, Italy}

\author{Giorgio Biasiol}
\affiliation{Laboratorio Nazionale TASC, CNR-INFM, AREA Science Park, Trieste I-34012 Trieste, Italy}

\author{Fabio Beltram}
\affiliation{NEST, Scuola Normale Superiore and CNR-INFM, Piazza dei Cavalieri 7, I-56126 Pisa, Italy}

\begin{abstract}
A surface-acoustic-wave (SAW) driven light-emitting-diode structure that can implement a single-photon-source for quantum-cryptography applications is demonstrated. Our lateral \textit{n-i-p} junction is realized starting from an undoped GaAs/AlGaAs quantum well by gating. It incorporates interdigitated transducers for SAW generation and lateral gates for current control. We demonstrate acoustoelectric transport and SAW-driven electroluminescence. The
acoustoelectric current can be controlled down to complete pinch-off by means of the lateral gates. 
\end{abstract}

\maketitle
Application of quantum physics to information processing and communication has opened the way to entirely innovative approaches to secure information sharing. From the pioneering work of Bennet and Brassard \cite{BB84}, several quantum-cryptographic protocols have been developed. Most of these are based on a quantum version of the classical one time-pad encription system in which the encription-decription key is shared through the exchange of quantum photonic states. Such an approach demands high repetition-rate streams of single photons and consequently fast single-photon sources (SPSs) and detectors. In this context a significant effort has been dedicated to the realization of quantum light sources that, generating a stream of pure single-photon state, will allow to overcome the security issues\cite{BrassardPRL2000} due to multi-photon emission of highly attenuated lasers (which are the most used SPS’s for cryptographic applications). Many physical quantum systems\cite{LounisRepProgPhys2005} have been proposed for the role of ideal quantum-cryptographic SPS and in particular solid-state devices appear to be very appealing thanks to their compatibility with standard industrial fabrication and telecommunication techniques. Some realisation of solid-state SPS have already been demonstrated and are mostly based on self-assembled, cavity-embedded quantum dots\cite{ShieldsNature2007,StraufNature2007}. An alternative approach, originally proposed by Foden \textit{et al.} \cite{FodenPRA2000}, consists in exploting surface-acoustic-wave (SAW) driven transport to transfer single-electrons from the \textit{n}-type to the \textit{p}-type region of a GaAs/AlGaAs heterostructure. A SAW-driven SPS is based on a lateral light-emitting diode with an intrinsic region interposed between the \textit{n} and the \textit{p} portions that hosts a constriction region capable of quantizing the acoustoelectric current to the single‑carrier regime already demonstrated for \textit{n-i-n} devices containing a 2-dimensional electron gas (2DEG)\cite{TalyanskiiPRB1997}. The main advantage of SAW-driven SPSs will be the high repetition rate (in the GHz range).

The main building block of our source is a SAW-driven lateral light-emitting diode\cite{CecchiniAPL2003,CecchiniAPL2004,CecchiniAPL2005,GellAPL2006,Gell2007}. These devices allowed the demonstration of SAW-driven light-emission, but the geometries explored so far are not compatible with the implementation of the tunable constriction region required for charge-transfer control down to the single-electron limit. Here we present a modified fabrication scheme for a SAW-driven and light-emitting \textit{n-i-p} lateral device which fills this gap. Our devices exploit the possibility of inducing by gating a two-dimensional layer of carriers in a quantum well (QW) contained in an undoped GaAs/AlGaAs heterostructure. In particular our design allows the formation of electron and hole gases in adjacent regions of the QW. Here we shall show that the electrical and emission properties of our devices can be controlled by means of SAWs. Furthermore we shall demonstrate an electrostatically tunable constriction channel that can control the charge transfer from the \textit{n} to the \textit{p} section of the device down to pinch-off.

\begin{figure}
\centering
\includegraphics[width=8.5cm]{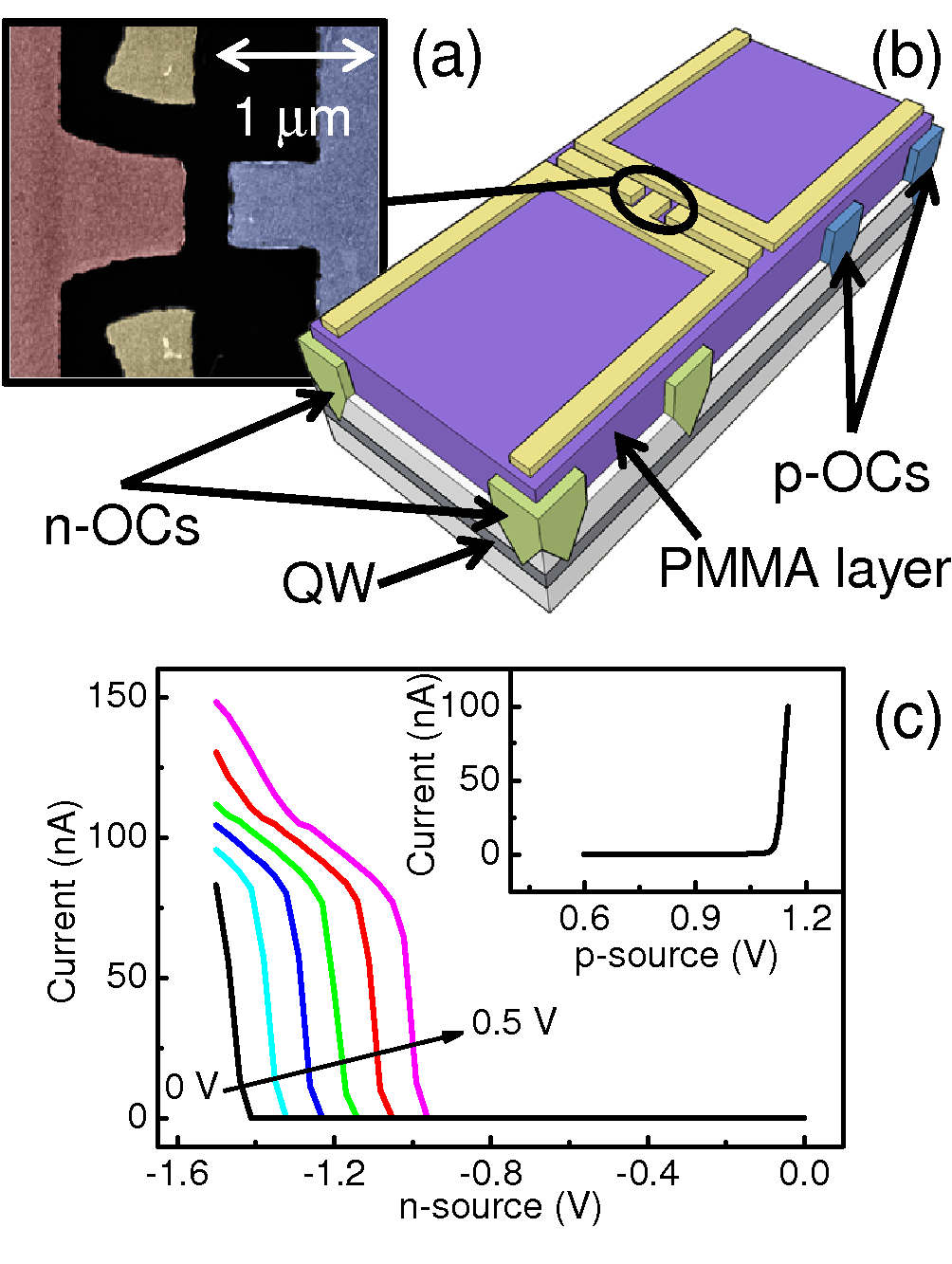}
\caption{(a) Scanning electron microscope image of the junction region. The image was colored to highlight diferrent regions: in red (blue) the \textit{n}-gate (\textit{p}-gate), in yellow the lateral gates. (b) Schematic picture of the device. (c) I/V characteristics of the \textit{n}-part of the device at a temperature of 5 K. Curves are taken at different voltages, from 0 V to 0.5 V, applied to the \textit{n}-gate. Inset: I/V characteristic of the \textit{p}-type region of the device. The \textit{p}-gate was kept grounded.} \label{Fig1}
\end{figure}

Our devices are fabricated on undoped GaAs/AlGaAs heterostructures containing a 20-nm wide GaAs QW located 40 nm below the surface. Mesas were defined by standard optical lithography and wet-etching procedures. \textit{n}-type Ni/AuGe/Au (10/180/100 nm) and \textit{p}-type Au/Zn/Au (5/60/100 nm) Ohmic contacts (OCs) were thermally-evaporated to locally provide donors and acceptors. As recently shown by 
Willet \textit{et al.}\cite{WilletAPL892007,WilletAPL912007}, it is possible to induce electrons or holes in an undoped QW by means of insulated metallic gates that partially overlap the OCs. In order to take advantage of this technique, our devices comprise a pair of Cr/Au (5/15 nm) gates: one of them partially overlaps the \textit{n}-OCs, the other one is partially overlapped to the \textit{p}-OCs. Insulation between OCs and gates was provided by a 300 nm-thick PMMA layer. These gates (in the following \textit{n}-gate and  \textit{p}-gate) make it possible to induce both electrons and holes in the QW by field effect. In this way a 2DEG and a two-dimensional hole gas (2DHG) are produced under the respective gates. Furthermore, by tuning the bias applied to the gates and to the OCs, it is possible to control the band profile across the \textit{n-i-p} lateral junction which is formed in the QW: in particular when the \textit{n}-gate and the \textit{p}-gate are biased at the same voltage no potential barriers are expected to arise along the entire length of the device. This results in a flat-band situation which is ideally suited to SAW-driven transport across the \textit{n-i-p} junction. Metallic layers in the region of propagation of the SAW were kept to a minimum by fabricating the gates with a 5-$\mu$m-wide U-shaped geometry and by choosing a gate thickness of 20 nm. This also minimizes the absorption of photons emitted following radiative electron-hole recombination. As shown in Fig.~\ref{Fig1} the central sections of the U-gates incorporate narrow ($\sim$600 nm) structures (NSs) which were designed to select a privileged area for \textit{n-i-p} conduction. A pair of lateral gates (LGs) are built  besides the NSs in the same processing steps of the U-gates. The NSs form the required constriction region and it can be electrostatically controlled by applying a voltage to the LGs. An interdigitated transducer (IDT), with a resonance frequency of 3 GHz, was fabricated to generate SAWs propagating from the \textit{n} to the \textit{p} side of the device.
Device electrical testing was performed by current-voltage characterization at 5 K. Measurements show that a 2DEG and a 2DHG are both induced by properly biasing the contacts and that it is possible to drive current across the \textit{n-i-p} junction. Figure~\ref{Fig1}c shows the current collected from one of the \textit{n}-type OC (the \textit{n}-drain) as a function of the bias applied to one of the other \textit{n}-type OC (the \textit{n}-source) when other OCs are left floating. Different curves correspond to different \textit{n}-gate voltages. Current flow between the two contacts demonstrates that a \textit{n}-type region was formed under the gated area of the device. Characteristics exhibit a threshold behavior corresponding to the alignment of the QW conduction-band bottoms in the \textit{n}-source and in the \textit{n}-gate regions. As expected, the \textit{n-i-n} threshold can be tuned by changing the voltage applied to the \textit{n}-gate. The same measurements were performed for the \textit{p}-part of the devices with similar results (see the inset of Fig.~\ref{Fig1}c). Leakage currents to the gates were negligible.
\begin{figure}[!ht]
\centering
\includegraphics[width=8.5cm]{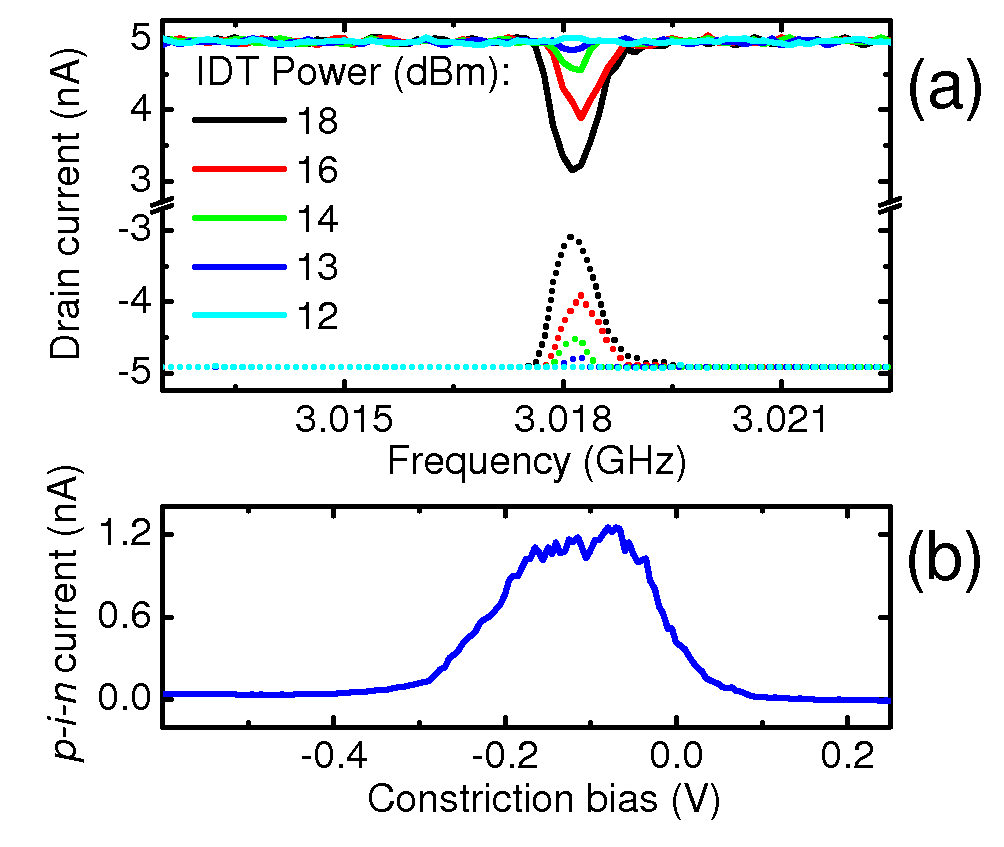}
\caption{(a) \textit{n}-drain (solid) and \textit{p}-drain (dashed) current as a function of the IDT excitation frequency at different excitation power levels when a fixed current of -5 nA (5 nA) was injected into the \textit{n} (\textit{p}) source. Out of resonance the carriers injected from the sources reach the drains of the same type. At resonance electron transport from the \textit{n} to the \textit{p} part is observed. (b) \textit{n-i-p} SAW driven transport as a funtion of the bias applied to the LGs. The IDT was excited at resonance (3.018 GHz) at a power of 18 dBm. Complete pinch-off of the current was observed both for positive and negative bias values.} \label{Fig2}
\end{figure}
Acoustoelectric transport properties of the devices were probed by taking advantage from the ability of the SAWs to extract electrons from a current flow \cite{CecchiniAPL2006}: we injected a constant electron current into the \textit{n}-source while measuring the current flowing out from the \textit{n}-drain. A 2DHG was induced in the \textit{p}-section of the device by injecting a current between \textit{p}-source and the \textit{p}-drain. The sources and the drains of the same type were chosen at opposite sides of the U-gates in order ensure carrier flow in the regions close to the NSs. Gates and QPC were kept grounded. Curves in the top  panel of Fig.~\ref{Fig2} show the current \cite{note} flowing into the \textit{p}-type (dashed lines) and \textit{n}-type (solid lines) drain as a function of the IDT excitation frequency and at different SAW power levels. When the IDT is excited out of resonance all the carriers injected into the sources are collected at the corresponding drain. At resonance (3.018 GHz), carrier transfer between the two regions is observed: a fraction of the injected electrons are steered by the SAW towards the \textit{p}-region, resulting in a negative peak in the \textit{n}-drain current. The electrons transferred into the 2DHG region recombine with holes resulting in a positive peak in the \textit{p}-drain current. Data reported in inset of Fig.~\ref{Fig3} show that the extraction efficiency depend on the \textit{n}-source current level, reaching the level of 40\% of steered electrons for a current of 1 $\mu$A at a SAW power level of 18 dBm.
\begin{figure}[!ht]
\centering
\includegraphics[width=8.5cm]{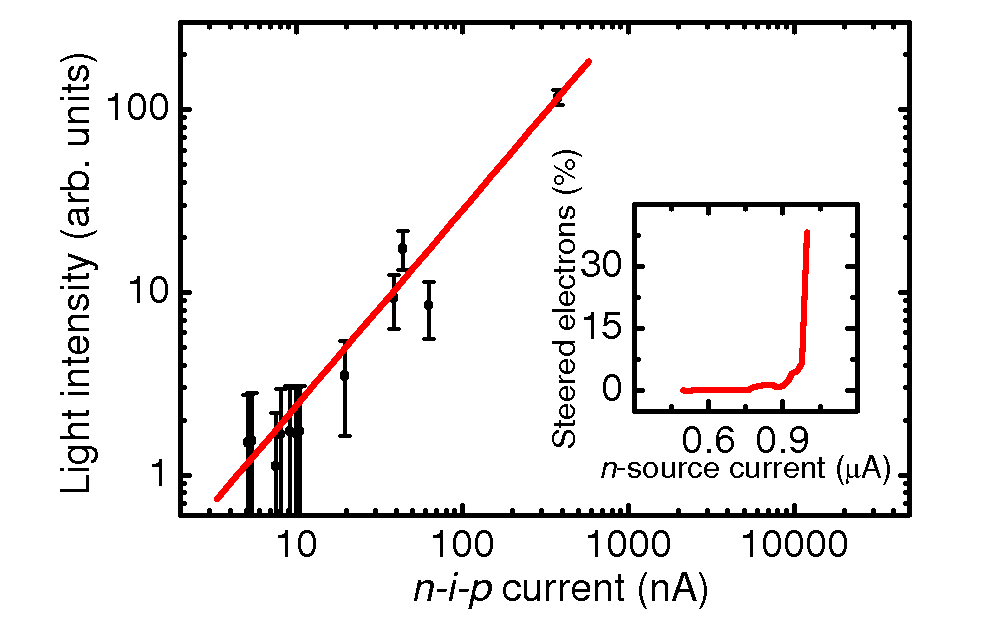}
\caption{Electroluminescence intensity as a function of the \textit{n-i-p} SAW-driven current. A linear dependence of the EL on the current was observed. Inset: fraction of electron steered toward the p-region as a function of the \textit{n}-source current. The IDT was excited at resonance (3.018 GHz) at a power of 18 dBm.} \label{Fig3}
\end{figure}

Figure~\ref{Fig2}b shows that the constriction can be narrowed down to complete acoustoelectric-current pinch-off by means of the LGs. This was observed both for positive and negative voltages. We argue that this behavior is due to the narrowing of the conduction channel for electrons at negative voltages as usually observed on 2DEGs, while at positive voltage the \textit{p}-NS region is depleted from holes preventing recombination. 

Corresponding to \textit{n-i-p} transport we observed an electroluminescence (EL) signal due to electron-hole radiative recombination. EL was focused on a cooled CCD camera by a high-numerical-aperture (0.68) aspherical lens placed inside the cryostat in order to maximize collection efficiency. This setup enabled us to obtain spatially-resolved snapshots of EL emission with a collection spot size of approximately 700 nm confirming that emission spots were located in the constriction region. Devices emitted for a broad range (from approximately 10 nA to almost 1 $\mu$A) of SAW-driven \textit{n-i-p} current values. Figure~\ref{Fig3} shows the dependence of EL intensity on SAW-driven \textit{n-i-p} current. The EL data plotted on a log-log chart as a function of the \textit{n-i-p} SAW-driven current follow a straight line with a 1.0$\pm$0.1 slope.

In conclusion we have shown a SAW-driven light-emitting junction using a geometry that allows SAW-driven gate-controlled transfer of electrons from the \textit{n}-type to the \textit{p}-type region of the device. We demonstrated acoustoelectric transport and SAW-driven electroluminescence. The acoustoelectric current can be controlled down to complete pinch-off by means of the lateral gates. Device optimization is currently being carried out to demonstrate single-photon operation. In particular, while for single-electron transport on a 2DEG the optimal geometries of the constriction region are well known, in our case we speculate that the design of the NSs must be further fine tuned owing to the fact that electron and hole gases are induced in an intrinsic QW.

This work was supported in part by the European Commission through the IP project SECOQC. G.~D.~S. wish to thank also the Fondazione Silvio Tronchetti Provera for financial support during this work.

\end{document}